\documentclass[aps,pra,preprint,groupedaddress]{revtex4}

\usepackage{graphicx}
\begin{document}

\title{Separability criterion for separate quantum systems}


\author{M.G. Raymer}
\email{raymer@oregon.uoregon.edu}
\author{A.C. Funk}
\affiliation{Oregon Center for Optics and Department of Physics, University of Oregon, Eugene,
Oregon, 97403, USA}

\author{B.C. Sanders}
\affiliation{Department of Physics and Centre for Advanced Computing-Algorithms and Cryptography, Macquarie University, Sydney, New South Wales, 2109 Australia}

\author{H. de Guise}
\affiliation{Department of Physics, Lakehead University, Thunder Bay, Ontario, P7B 5E1 Canada}


\date{\today}

\begin{abstract}
 Entanglement, or quantum inseparability, is a crucial resource in quantum information 
applications, and therefore the experimental generation of separated yet entangled systems is of 
paramount importance.  Experimental demonstrations of inseparability with light are not uncommon, but 
such demonstrations in physically well-separated massive systems, such as distinct gases of atoms, are 
new and present significant challenges and opportunities. Rigorous theoretical criteria are needed for 
demonstrating that given data are sufficient to confirm entanglement. Such  criteria for experimental data 
have been derived for the case of continuous-variable systems obeying the Heisenberg-Weyl (position-
momentum) commutator. To address the question of experimental verification more generally, we 
develop a sufficiency criterion for arbitrary states of two arbitrary systems. When applied to the recent 
study by Julsgaard, Kozhekin, and Polzik [Nature \textbf{413}, 400 - 403 (2001)] of spin-state entanglement of 
two separate, macroscopic samples of atoms, our new criterion confirms the presence of spin 
entanglement. 
\end{abstract}

\maketitle

Entanglement, or quantum inseparability, is a profound property of nature that enables  
information to be stored, communicated, and processed in a decidedly non-classical fashion.\cite{ref:Niel_Chuang}
Entanglement has long been observed in the states of small numbers of microscopic objects such as 
electrons or photons. Only recently have there been efforts to create and observe entanglement in the state 
of massive macroscopic objects, such as the collective spins of two separate atomic vapors.\cite{ref:JKP,ref:Bigelow} It is 
important therefore to develop a sufficient criterion, which, if satisfied, would unambiguously verify that 
an experiment has displayed entanglement.\par

Previous significant work has been done to find a sufficiency criterion that is valid for 
continuous-variable systems obeying the Heisenberg-Weyl (HW) commutator, valid for position-
momentum variables and, similarly, for light-field amplitudes. \cite{ref:Duan,ref:Simon}  Such a condition is not strictly 
valid, however, for collective spin systems, although an approximate correspondence was proposed for 
certain special spin states and used to analyse a recent experiment by Julsgaard, Kozhekin, and Polzik 
(JKP).\cite{ref:JKP} This study was aimed at demonstrating spin-state entanglement for two separate, macroscopic 
samples of atoms containing around $10^{12}$ atoms each.  We derive a sufficiency condition for the existence 
of entanglement between two arbitrary quantum systems, including spin systems, in pure or mixed states. 
This allows us, for example, to confirm rigorously the presence of entanglement  in the experiment of 
JKP. This new criterion is general, and so may find application in other experimental studies. \par

Two distinct quantum systems 1 and 2 are said to be entangled if their joint density operator $\hat{\rho}$  is 
inseparable, that is, if  $\hat{\rho}$ cannot be represented as a convex sum of density operators $\hat{\rho}_{1i}$  and  $\hat{\rho}_{2i}$  for the 
two physically separated systems, \cite{ref:Werner,ref:Lewenstein}
\begin{eqnarray}
\hat{\rho}&=&\sum_{i} p_{i}\hat{ \rho}_{1i}\otimes \hat{\rho}_{2i},
\label{eq:manu1}
\end{eqnarray}
with $p_{i}$  a set of non-negative, normalized probabilities. If their joint state is separable (not entangled), 
then it must be possible to express the density operator in the form Eq. (\ref{eq:manu1}). One physical interpretation of 
entanglement is that it represents a correlation between two systems that is stronger than can exist in any 
classical (local, realistic) theory.\cite{ref:Bell}  \par

A convincing demonstration of entanglement would prove a violation of the separability 
condition Eq. (\ref{eq:manu1}). In attempting to demonstrate inseparability between the spin variables of two separated 
atomic samples, JKP employ non-local Bell measurements on the spin variables and relate these spin 
variables to canonical position and momentum operators obeying the Heisenberg-Weyl (HW) 
commutator $[\hat{q}_{j},\hat{p}_{k}]=i\delta_{jk}$ ($j,k=1,2$).  By establishing this approximate correspondence, JKP then 
adapt a criterion by Duan \textit{et al.} \cite{ref:Duan} and by Simon \cite{ref:Simon}, which applies to coupled oscillators (and 
specifically to squeezed light). The ``HW'' criterion that is sufficient for inseparability is \cite{ref:Duan,ref:Simon}
\begin{eqnarray}
\text{var}(\hat{q}_{1}+\hat{q}_{2})+\text{var}(\hat{p}_{1}-\hat{p}_{2})\geq2,
\label{eq:manu2}
\end{eqnarray}
where $\text{var}(... )$ represents the statistical variance. JKP's criterion is an expression analogous to Eq.(\ref{eq:manu2}), 
predicated on the assumption that for certain states spin operators can be approximately replaced by 
canonical position and momentum operators. \par

Although the shortcut proposed by JKP offers an appealing connection between criteria for 
demonstrating entanglement in squeezed-light systems and in spin ensembles, the validity of this 
correspondence is far from obvious, and can lead to misconceptions regarding transformations between 
different bases that are quite distinct from the Fourier transform nature of the canonical position-
momentum transformations. Before returning to a consideration of entanglement in collective spin 
systems, we first establish a criterion for inseparability that is applicable to any algebra, including that for 
spin. We do this by generalizing the calculations of Duan et al \cite{ref:Duan} and of Berry and Sanders \cite{ref:Berry}. \par

We consider two systems 1 and 2, and two observables for each,  $\hat{A}_{1}, \hat{B}_{1}$ for system 1 and $\hat{A}_{2}, \hat{B}_{2}$  
for system 2, that obey $[\hat{A}_{i},\hat{B}_{j}]=\delta_{i,j}\hat{C}_{j}$ . Define linear combinations,
\begin{eqnarray}
\hat{u}&=&\alpha\hat{A}_{1}+\beta\hat{A}_{2} \nonumber \\
\hat{v}&=&\alpha\hat{B}_{1}-\beta\hat{B}_{2},
\label{eq:manu3}
\end{eqnarray}
for $\alpha, \beta$  arbitrary real coefficients. Equation (\ref{eq:manu1}) implies for the variance 
\begin{eqnarray}
\text{var}(\hat{u})&=&\sum_{i}p_{i}[ \alpha^{2}\langle (\Delta\hat{A}_{1})^{2}\rangle_{i} + \beta^{2}\langle (\Delta\hat{A}_{2})^{2}\rangle_{i} ] \nonumber \\
&&+S,
\label{eq:manu4}
\end{eqnarray}
where $\Delta\hat{A}_{k}=\hat{A}_{k}-\langle\hat{A}_{k}\rangle_{\rho}$   and $\langle...\rangle_{\rho}$  denotes an average over $\hat{\rho}$.  The quantity $S$ is $S=\sum_{i}p_{i}\langle\hat{u}\rangle_{i}^{2}-\left(\sum_{i}p_{i}\langle\hat{u}\rangle_{i}\right)^{2}$, where $\langle...\rangle_{i}$  denotes the average over the product density operator $\hat{\rho}_{1i}\otimes\hat{\rho}_{2i}$.  The Schwarz inequality implies in general that  $S\geq0$. Doing the same for $\hat{v}$ and adding the 
results gives
\begin{eqnarray}
\text{var}(\hat{u})+\text{var}(\hat{v})&\geq&\sum( p_{i}[ \alpha^{2}\langle (\Delta\hat{A}_{1})^{2}\rangle_{i} + \beta^{2}\langle (\Delta\hat{A}_{2})^{2}\rangle_{i} ] + \nonumber \\
&& p_{i}[ \alpha^{2}\langle (\Delta\hat{B}_{1})^{2}\rangle_{i} + \beta^{2}\langle (\Delta\hat{B}_{2})^{2}\rangle_{i} ]),
\label{eq:manu5}
\end{eqnarray}
or
\begin{eqnarray}
\text{var}(\hat{u})+\text{var}(\hat{v})&\geq&\alpha^{2}[ \langle (\Delta\hat{A}_{1})^{2}\rangle_{\rho} + \langle (\Delta\hat{B}_{1})^{2}\rangle_{\rho} ]+\nonumber \\
&&  \beta^{2}[ \langle (\Delta\hat{A}_{2})^{2}\rangle_{\rho} +\langle (\Delta\hat{B}_{2})^{2}\rangle_{\rho} ].
\label{eq:manu6}
\end{eqnarray}

Equation (\ref{eq:manu6}) is always satisfied for any separable state, with respect to any variables (discrete or 
continuous) belonging to any algebra. If one can measure all the corresponding quantities and find a 
violation of Eq. (\ref{eq:manu6}), then one demonstrates that the state is inseparable.  \par

	The general commutator  $[\hat{A}_{i},\hat{B}_{j}]=\delta_{ij}\hat{C}_{j}$  implies the uncertainty relation $\Delta A_{i} \Delta B_{i} \geq (1/2) C_{i},$  
where $C_{i}=|\langle \hat{C}_{i}\rangle|=|\text{Tr}(\hat{\rho}[\hat{A}_{i},\hat{B}_{i}])|,$  ($i=1,2$). This implies the less restrictive relation  $\Delta A_{i}^{2} + \Delta B_{i}^{2}\geq C_{i}$, with equality only for $\Delta A_{i}^{2} =C_{i}/2$.  Inserting this into Eq. (\ref{eq:manu6}) gives, for any separable state, 
\begin{eqnarray}
\text{var}(\hat{u})+\text{var}(\hat{v})&\geq&\alpha^{2}C_{1} +\beta^{2}C_{2}.
\label{eq:manu7}
\end{eqnarray}
This is our main result. A related criterion has been recently found for the case of pure states of spin 
systems. \cite{ref:Berry} In the special case  $\alpha=\beta=1,$ Eq. (\ref{eq:manu7}) gives
\begin{eqnarray}
\text{var}(\hat{A}_{1}+\hat{A}_{2})+\text{var}(\hat{B}_{1}-\hat{B}_{2})&\geq&C_{1} +C_{2}.
\label{eq:manu8}
\end{eqnarray}
\par

Equation (\ref{eq:manu7}) is not a tight bound. That is, it is necessary for any separable state to satisfy Eq. (\ref{eq:manu7}), 
but it need not be violated for every entangled (i.e., inseparable) state. So Eq. (\ref{eq:manu7}) is a necessary but not 
sufficient condition for separability. A sufficient \emph{and} necessary criterion that is experimentally accessible 
for spin ensembles is not known. For the special case of Gaussian states of Heisenberg-Weyl systems 
with $C_{1}=C_{2}=1$, Eq.(\ref{eq:manu8}) reduces to Eq. (\ref{eq:manu2}), which has been shown by Duan \textit{et al.} \cite{ref:Duan} and Simon \cite{ref:Simon} to be 
a sufficient \emph{and necessary} condition for separability in this case. \par

	In the JKP study, the variables of interest are the projections  $\hat{J}_{x}, \hat{J}_{y}, \hat{J}_{z}$ of the collective  spins of 
two atomic samples, 1 and 2.  The experiment \cite{ref:JKP} can be analyzed by choosing $\hat{A}_{1}=\hat{J}_{y1}, \hat{B}_{1}=\hat{J}_{z1},  \hat{A}_{2}=\hat{J}_{y2}, \hat{B}_{2}=-\hat{J}_{z2}$.  Then $C_{1}+C_{2}=|\langle \hat{J}_{x1}\rangle| +|\langle-\hat{J}_{x2}\rangle | = 2 |\langle \hat{J}_{x1}\rangle |$, and separability requires, from 
Eq. (\ref{eq:manu8}),
\begin{eqnarray}
\text{var}(\hat{J}_{y1}+\hat{J}_{y2})+\text{var}(\hat{J}_{z1}+\hat{J}_{z2})&\geq&2|\langle \hat{J}_{x1} \rangle |.
\label{eq:manu9}
\end{eqnarray}
Equation (\ref{eq:manu9}) yields a rigorous criterion: if this inequality is violated, then entanglement has been 
demonstrated.\par

This result is similar in form to JKP's Eq. (1), reviewed below, but is distinct in several important 
respects. The first is that our criterion for demonstrating inseparability is expressed entirely in terms of 
the spin operators and does not entail any approximations. This result is valid even without the restriction 
that a large number of atoms is required. There is no recourse, nor any need for recourse, to canonical 
position and momentum operators or to the criterion for squeezed oscillators. The second difference is 
that Eq. (\ref{eq:manu9}) is a valid criterion for arbitrary states Ð not only for certain extremum states as in the criterion 
of JKP. The final difference is that the right-hand side of Eq. (\ref{eq:manu9}) is the expectation value with respect to 
the state under investigation, rather than being determined by a quantity defined in terms of some 
``classical'' value. \par

The present result puts on a firm theoretical ground the criterion used by JKP as a \emph{necessary}  
criterion for separability. The violation of Eq. (\ref{eq:manu9}) by the data in the JKP study can be taken as an 
indication of the breakdown of separability.  Nevertheless, the question of an experimentally accessible, 
\emph{sufficient} condition, even for special classes of states (e.g., the Gaussian ones for the case of HW 
systems), is still an open one for the case of spin systems. \par

Here we present arguments that one cannot take the approximate correspondence between spin 
variables and HW variables too literally, as it can lead to errors if care is not taken. (Our approach avoids 
the problematic extrapolation of HW results.) For example, large errors occur when calculating a change 
of basis if one uses eigenstates of $\hat{J}_{y}$ and $\hat{J}_{z}$ as basis states and assumes that these transform 
approximately as HW variables do. These errors persist even for the extremum states considered by JKP.\par

To review JKP's analysis, the collective-spin vector operator $\hat{J}$ (total angular momentum) of a 
collection of $N$ atoms (where $N$ may be known or statistically distributed) is defined to have $x$-component 
 $\hat{J}_{x}=\sum_{i=1}^{N}\hat{J}_{x}^{(i)}$, and similarly for $\hat{J}_{y}$ and $\hat{J}_{z}$.  These components obey the algebra  $[\hat{J}_{y},\hat{J}_{z}]=i\hat{J}_{x}$, et. cycl., and commute with $\hat{J}^{2}$; the number of atoms $N$ determines the corresponding irreducible representation. 
For $N=2j$, one choice for an orthonormal basis comprises $|j,m_{x}\rangle_{x}$ which satisfy the eigenvalue 
relations  $\hat{J}^{2}|j,m_{x}\rangle_{x}=j(j+1)|j,m_{x}\rangle_{x}$ and $\hat{J}_{x}|j,m_{x}\rangle_{x}=m_{x}|j,m_{x}\rangle_{x}$.  For $\overline{J}_{x}$ equal to some ``large 
classical'' real number ($ > 0$), JKP define operators $\hat{Q}=\hat{J}_{y}/\sqrt{\overline{J}_{x}}$ and $\hat{P}=\hat{J}_{z}/\sqrt{\overline{J}_{x}}$ satisfying $[\hat{Q},\hat{P}]=i\hat{J}_{x}/\overline{J}_{x}$ \cite{ref:Note}.\par

\begin{figure}
\includegraphics{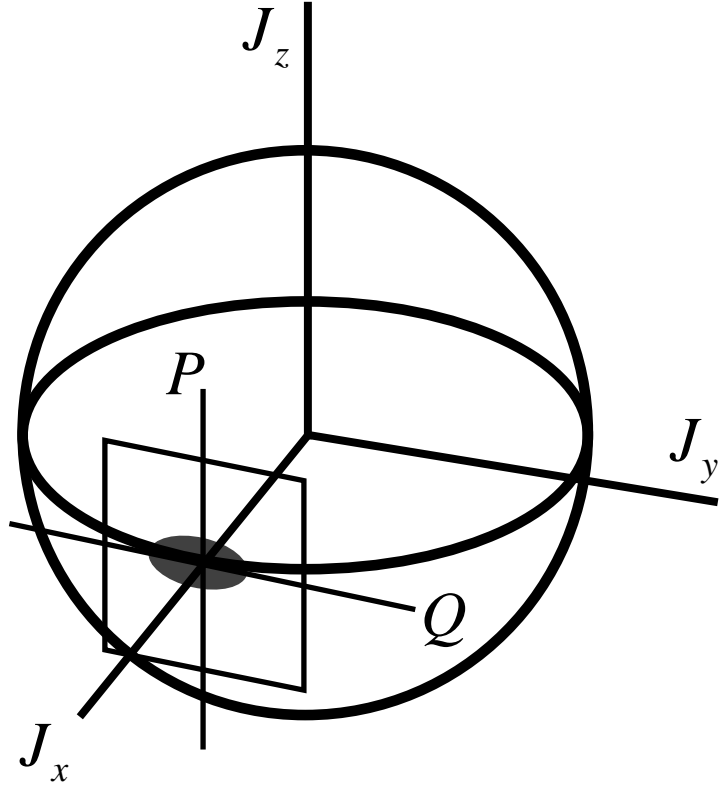}
\caption{``Extremum'' angular-momentum states having large total $J$ and total $J_{x}$$\widetilde{ = }$$J$  can be visualized as occupying the shaded region tightly concentrated near the $J_{x}$ `pole' in a space with axes $J_{x}$, $J_{y}$, $J_{z}$.  The quasi-continuous variables $Q$ and $P$ can be thought of as forming approximately the Cartesian coordinates  of the tangent plane touching the sphere with radius $J$.   }
\label{fig:poinc}
\end{figure}

Consider extremum states $|\Psi\rangle$ having narrow support over approximately equal values of $m_{x}\cong\overline{J}_{x}$, where $\overline{J}_{x}$ is a large, state-independent real number. Such extremum states  can be visualized as 
tightly concentrated near the $J_{x}$ `pole' in a space with axes  $J_{x}, J_{y}, J_{z}$, as illustrated in Fig. \ref{fig:poinc}.  JKP 
suggest that for such states one can approximate $\hat{J}_{x}/\overline{J}_{x}$ by the unity operator to obtain $[\hat{Q},\hat{P}]=i$. This 
commutator, along with Eq.(\ref{eq:manu2}), would lead directly to the \emph{necessary}  criterion for separability  in the form 
of JKP's Eq. (1). This result is correct in a restricted sense, as noted above.  \par

Nevertheless, there are difficulties with taking this approximate approach too literally. This is 
evidenced by the fact that a basis transformation between the eigenstates of $\hat{Q}=\hat{J}_{y}/\sqrt{\overline{J}_{x}}$ and 
 $\hat{P}=\hat{J}_{z}/\sqrt{\overline{J}_{x}}$ is not given by a Fourier transform, despite the commutator between the operators being 
forced to be a constant, which seemingly implies that the eigenstates have overlap $\langle P|Q\rangle\propto \exp(-i P Q)$. 
The inapplicability of the Fourier transform is apparent by attempting this transformation. In the $J_{y}$ basis,
\begin{eqnarray}
|\Psi\rangle &=& \sum_{2j=0}^{\infty}  \sum_{m_{y}=-j}^{j}|j,m_{y}\rangle_{y}   \hspace{.1cm}_{y}\langle j,m_{y}|\Psi\rangle \nonumber \\
&= &\sum_{2j=0}^{\infty}\sum_{m_{y}=-j}^{j}|j,m_{y}\rangle_{y} C_{y}(j,m_{y}),
\label{eq:manu10}
\end{eqnarray}
and the summation notation means sum $j$ over nonnegative half-integers. For the extremum  states, with 
large mean-$j$ value $\overline{j}$ (say $10^{12}$), the coefficients $C_{y}(j,m_{y})$ are non-negligible only in the vicinity of $m_{y}=0$. \par

The same state represented in the $J_{z}$ basis is
\begin{eqnarray}
|\Psi\rangle &=& \sum_{2j=0}^{\infty}\sum_{m_{z}=-j}^{j} C_{z}(j,m_{z}) |j,m_{z}\rangle_{z}.
\label{eq:manu11}
\end{eqnarray}
The two sets of coefficients are related by 
\begin{eqnarray}
C_{z}(j,m_{z})&=&\sum_{m_{y}=-j}^{j}\hspace{0.1cm} _{z}\langle j,m_{z}|j,m_{y}\rangle_{y} C_{y}(j,m_{y}),
\label{eq:manu12}
\end{eqnarray}
Angular momentum algebra gives $_{z}\langle j,m_{z}|j,m_{y}\rangle_{y}=d_{m_{z},m_{y}}^{j}(\pi/2)$, where the elements of the rotation 
matrix (reduced Wigner function) are \cite{ref:Varsh}
\begin{eqnarray}
d_{mm'}^{j}(\pi/2)&=&2^{-j}\sqrt{\frac{(j+m')!(j-m')!}{(j+m)!(j-m)!}}\sum_{k=0}^{j-m}{{j+m}\choose{j+m'-k}}{ {j-m}\choose{k}} (-1)^{m-m'+k}
\label{eq:manu13}
\end{eqnarray}
The basis transformation Eq.(\ref{eq:manu12}) is entirely different from a Fourier transformation, in which the  $m_{y}, m_{z}$
values would be replaced by quasi-continuous variables $m_{y}\rightarrow J_{y}=Q\sqrt{\overline{J}_{x}}$, $m_{z}\rightarrow J_{z}=P\sqrt{\overline{J}_{z}}$, and the 
transformation would be 
\begin{eqnarray}
C_{z}(P)&=&\int_{-\infty}^{\infty} dQ (1/2\pi)^{1/2} \exp(-iPQ) C_{y}(Q).
\label{eq:manu14}
\end{eqnarray}
The asymptotic form of the reduced Wigner function Eq.(\ref{eq:manu13}) is given in the Appendix, where it is seen 
not to be approximated by the Fourier transform kernel. Furthermore, these transformations differ in a 
qualitative way: Whereas  the transformation kernel of Eq.(\ref{eq:manu14}) is necessarily complex, there exists a 
choice of phase that makes the correct kernel real, as in Eq. (\ref{eq:manu13}).\par

A concrete example, given in the Appendix, illustrates the large errors that can result from using 
the Fourier transform. There we consider a specific state satisfying the assumed extremum properties 
($\overline{J}_{y},\overline{J}_{z}\ll\overline{J}_{x}$), which would presumably make the commutator $[\hat{Q},\hat{P}]=i$ approximately correct. Upon 
making a basis change from the $J_{y}$ basis to the $J_{z}$ basis, we find, using the correct Eq. (\ref{eq:manu12}), that the mean 
value of $\hat{J}_{z}$ is given by a formula consistent with $\overline{J}_{z}\ll\overline{J}_{x}$. However, when (provisionally) using the 
Fourier transform for the basis change calculation we compute a mean value $\overline{J}_{z}=-(\pi/2)\overline{J}_{x}$. This is 
incorrect, as for this state $\overline{J}_{z}$ must be much smaller than $\overline{J}_{x}$.  This demonstrates the complete breakdown 
of a simple, direct replacement of the spin-operator algebra by the HW algebra, leading to the need for the 
more careful derivation we provided in the first half of this paper.\par

In conclusion, Eq. (\ref{eq:manu7}) provides a necessary condition for separability for arbitrary states of two 
general systems. This condition is accessible to experimental tests in that it involves measurements of 
only several low-order moments. When applied to collective angular-momentum variables in macroscopic 
atomic systems, the new criterion confirms the one used by JKP in their experimental study.\cite{ref:JKP} The 
problem of finding sufficient conditions for special classes of angular-momentum states remains to be 
solved. \par

The complete replacement, for all purposes, of the collective angular-momentum algebra by the 
simpler HW (position-momentum) algebra is not valid, even for extremum states that are nearly confined 
to a small region in angular- momentum space, corresponding to highly polarized atomic samples. We do 
not intend to imply that the use of the approximate commutator $[\hat{Q},\hat{P}]=i$ will always lead to large errors. 
If one evaluates operator moments involving only states confined to the proper extremum region, then 
only small errors are incurred, as is well known. We caution, however, that one cannot assume the 
validity of state expansions in basis states having the same properties as $Q$ and $P$ eigenstates.  \par

Finally, it is interesting to address the question - what states, if any, are conjugate to the $|j,m\rangle_{y}$ states through a Fourier transformation? The answer is the SU(2) phase states. In the SU(2) phase 
formulation \cite{ref:Vourdas} one constructs the $(2j+1)$-dimensional basis from phase states, defined as $|j,\theta_{k}\rangle_{y}=(2j+1)^{-1/2}\sum_{m=-j}^{j}e^{im\theta_{k}}|j,m\rangle_{y}$, with $\theta_{k}=k\pi/(2j+1)$.  From this we obtain the desired Fourier 
transform kernel $_{y}\langle j,\theta |j,m\rangle_{y}=e^{im\theta}/\sqrt{2j+1}$.  Even though the basis change from $|j,m\rangle_{y}$ states to 
phase states is a (discrete) Fourier transformation, the phase operators $\hat{\phi}_{y}$ constructed for this 
representation do \emph{not} naturally yield a commutator $[\hat{J}_{y},\hat{\phi}_{y}]=i$.  Therefore there is not an exact way to use 
this correspondence to construct an equivalent HW algebra.

\section{Appendix}
When $j$ is large and $m,m'\ll j$ the reduced Wigner function Eq. (\ref{eq:manu13}) is well approximated by 
using Stirling's formula  to give 
\begin{eqnarray}
d_{mm'}^{j}(\pi/2)&\cong&\sqrt{\frac{2}{\pi j}}\exp(+|m^{2}-m'^{2}|/2j)\cos\left((j+m-m')\frac{\pi}{2}\right).
\label{eq:manu15}
\end{eqnarray}
This does not approximate to the Fourier transform kernel.\par

As an illustration of the large errors that can arise when using the Fourier transform to execute a 
basis change between $J_{y}$ and $J_{z}$ bases, consider the state with
\begin{eqnarray}
C_{y}(j,m_{y})&=&\frac{\exp(-\overline{j})\alpha_{1}^{j+m_{y}}\alpha_{2}^{j-m_{y}}}  {\sqrt{(j+m_{y})!(j-m_{y})!}},
\label{eq:manu16}
\end{eqnarray}
where $\alpha_{k}=|\alpha_{k}|\exp(i\phi_{k})$.  The mean values for this state are $\overline{j}=(|\alpha_{1}|^{2}+|\alpha_{2}|^{2})/2$, $\overline{m}_{y}=\overline{J}_{y}=(|\alpha_{1}|^{2}-|\alpha_{2}|^{2})/2$, $\overline{J}_{z}=|\alpha_{1}\alpha_{2}|\cos(\phi_{2}-\phi_{1})$, and $\overline{J}_{x}=|\alpha_{1}\alpha_{2}|\sin(\phi_{2}-\phi_{1})$. We consider 
states such that $\phi_{2}-\phi_{1}\cong\pi/2$ and $|\alpha_{1}|^{2}-|\alpha_{2}|^{2}\ll|\alpha_{1}|^{2}+|\alpha_{2}|^{2}$, and hence are in the considered 
extremum class, with $\overline{J}_{y},\overline{J}_{z}\ll \overline{J}_{x}$. \par

In the $J_{z}$ basis this same state is represented exactly by (using Eqs. (\ref{eq:manu12},\ref{eq:manu13}))
\begin{eqnarray}
C_{z}(j,m_{z})&=&\frac{\exp(-\overline{j})\beta_{1}^{j+m_{z}}\beta_{2}^{j-m_{z}}}{\sqrt{(j+m_{z})!(j-m_{z})!}},
\label{eq:manu17}
\end{eqnarray}
with $\beta_{1}=(\alpha_{2}+\alpha_{1})/\sqrt{2}$ and $\beta_{2}=(\alpha_{2}-\alpha_{1})/\sqrt{2}$. The mean values $\overline{J}_{x}, \overline{J}_{y}, \overline{J}_{z}$ are unchanged by the 
change of basis, but we now have $\overline{m}_{z}=\overline{J}_{z}=|\alpha_{1}\alpha_{2}|\cos(\phi_{2}-\phi_{1})$.\par

How does this exact result compare with that obtained by assuming that the HW commutator is 
valid, which requires that we transform Eq. (\ref{eq:manu16}) by the Fourier relation? To carry this out we first find an 
accurate approximation to Eq. (\ref{eq:manu16}), using Stirling's formula, which gives, for $j\cong\overline{j}$ large (e.g. $10^{12}$) and $m_{y}, \overline{m}_{y}\ll\overline{j}$
 (and arbitrary phases $\phi_{1}, \phi_{2}$), 
\begin{eqnarray}
C_{y}(j,m_{y})&\cong &C(j) (\pi \overline{j})^{-1/4} \exp[ -(m_{y}-\overline{m}_{y})^{2}/{2 \overline{j}}]\exp[i(\phi_{1}-\phi_{2})m_{y}],
\label{eq:manu18}
\end{eqnarray}
with $C(j)=\sqrt{\exp(-2\overline{j})(2\overline{j})^{2j}/(2j)!}\exp[i(\phi_{1}+\phi_{2})j]$, which is a relatively narrow function of $j$. \par

Using Eq. (\ref{eq:manu14}) to transform Eq. (\ref{eq:manu18}) we find (with the provisional result indicated by the tilde)
\begin{eqnarray}
\widetilde{C}_{z}(j,m_{z})&\cong&C(j)\left( \frac{\overline{j}}{\pi \overline{J}_{x}^{2}} \right)^{1/4}\exp\left[\frac{-(m_{z}-\widetilde{m}_{z})^{2}}{2(\overline{J}_{x}^{2}/\overline{j})} \right]
\exp\left[ -i\left( \frac{m_{z}\overline{m}_{y}}{\overline{J}_{x}} -(\phi_{1}-\phi_{2})\overline{m}_{y} \right)  \right],
\label{eq:manu19}
\end{eqnarray}
which means that the mean value of $\hat{J}_{z}$ is given provisionally by $\overline{J}_{z}=\widetilde{m}_{z}=(\phi_{1}-\phi_{2})\overline{J}_{x}$.  This predicted 
value for $\overline{J}_{z}$ is quite incorrect. As an example consider $\phi_{2}-\phi_{1}=\pi/2$.  Equation (\ref{eq:manu17}) predicts (correctly 
and exactly) that $\overline{J}_{z}$ has a mean value $\overline{J}_{z}=0$, while Eq. (\ref{eq:manu19}) predicts a mean value $\overline{J}_{z}=\widetilde{m}_{z}=-(\pi/2)\overline{J}_{x}$.  This is incorrect, as for this state $\overline{J}_{z}$ must be much smaller than $\overline{J}_{x}$. \par

For completeness, the correct Eq. (\ref{eq:manu17}) for the state in the $J_{z}$ basis can be well approximated 
using Stirling's formula, (since $|\beta_{1}|^{2}-|\beta_{2}|^{2}\ll|\beta_{1}|^{2}+|\beta_{2}|^{2}$), giving
\begin{eqnarray}
\widetilde{C}_{z}(j,m_{z})&\cong&C(j)\exp[-(m_{z}-\overline{m}_{z})^{2}/2\overline{j}]\exp[i(\phi_{1}'-\phi_{2}')m_{z}],
\label{eq:manu20}
\end{eqnarray}
where $\phi_{k}'=\arg[\beta_{k}]$.  Equations (\ref{eq:manu20}) and (\ref{eq:manu19}) differ in two important ways - both in the phase structure 
and in the predicted mean value of  $m_{z}$.

\begin{acknowledgments}
 We wish to thank Stephen Bartlett and Dominic Berry for helpful discussions, and 
Peter Braun for a tip on arriving at Eq. (15). Support for this project has been provided by the National 
Science Foundation grants PHY-0140370, ECS-9977127, the Army Research Office grant DAAD19-99-
1-0344, an Australian Research Council International Research Exchange Scheme grant, and an 
Australian Research Council Large Grant. 
\end{acknowledgments}


\begin{thebibliography}{12}
\bibitem[{ref()}]{ref:Note}
\bibinfo{note}{We have rewritten the operators and commutator using the usual
  right-hand coordinate system convention, rather than the left-handed system
  used by JKP}.
\end{thebibliography}
\end{document}